\documentstyle[aps, epsfig]{revtex}

\def\xslash#1{{\rlap{$#1$}/}}
\def\half{\frac{1}{2}}
\def\beq{\begin{equation}}
\def\eeq{\end{equation}}
\def\beqa{\begin{eqnarray}}
\def\eeqa{\end{eqnarray}}
\def\iar{\begin{array}{l}}
\def\ear{\end{array}}

\begin{document}
\draft
\title{Renormalization of the Cabibbo-Kobayashi-Maskawa Quark Mixing Matrix}
\author{Yong Zhou$^{a}$}
\address{$^a$ Institute of Theoretical Physics, Chinese Academy of Sciences, 
         P.O. Box 2735, Beijing 100080, China}
\maketitle

\begin{abstract}
We have investigated the present renormalization prescriptions of 
Cabibbo-Kobayashi-Maskawa (CKM) matrix. When considering the prescription which is 
formulated with reference to the case of zero mixing we find it doesn't satisfy 
the unitary condition of the bare CKM matrix. After added a delicate patch this
problem can be solved at one-loop level. 
In this paper We generalize this prescription to all loop levels and keep the 
unitarity of the bare CKM matrix, simultaneously make the amplitude of an arbitrary 
physical process involving quark mixing convergent and gauge independent. We also 
find that in order to keep the CKM counterterms gauge independent the unitarity of
the bare CKM matrix must be preserved.
\end{abstract}


\section{Introduction}

Since the exact examination of the Cabibbo-Kobayashi-Maskawa (CKM) quark mixing 
matrix \cite{c1,c2,c3,c4,c5,c6} has been developed quickly, the renormalization 
of CKM matrix becomes very important. This was realized for the Cabibbo angle in 
the standard model (SM) with two fermion generations in a pioneering paper by 
Marciano and Sirlin \cite{c7} and for the CKM matrix of the three-generation SM 
by Denner and Sack \cite{c8} more than a decade ago. In recent years many people 
have discussed this issue \cite{c9,c10,c11}, but a completely self-consistent 
scheme to all loop levels has been not obtained. In this paper we try to solve this 
problem and give some instructive conclusion.

In general, a CKM matrix renormalization prescription needs to satisfy three 
criterions, as Diener has declared \cite{c12}:
\begin{enumerate}
\item In order to make the transition amplitude of any physical process involving
quark mixing ultraviolet finite, the CKM counterterm must cancel out the 
ultraviolet divergence left in the loop-corrected amplitudes. On the other hand it
must include proper infrared divergence for the sake of infrared finiteness of the
final scattering cross-section including soft quanta emission.
\item It must guarantee the transition amplitude of any physical process involving
quark mixing gauge parameter independent \cite{c13}, which is a fundamental 
requirement. 
\item SM requires the bare CKM matrix $V^0$ is unitary,

\beq
\sum_k V^0_{ik} V^{0\ast}_{jk}\,=\,\delta_{ij}
\eeq
with $i,j,k$ the generation index and $\delta_{ij}$ the unit matrix element. 
If we split the bare CKM matrix element into the renormalized one and its 
counterterm 

\beq
  V^0_{ij}=V_{ij}+\delta V_{ij}
\eeq
and keep the unitarity of the renormalized CKM matrix, the unitarity of the bare 
CKM matrix requires 

\beq
  \sum_k (\delta V_{ik}V^{\ast}_{jk}+V_{ik}\delta V^{\ast}_{jk}+
  \delta V_{ik}\delta V^{\ast}_{jk})\,=\,0
\eeq
\end{enumerate}

Until now there are many papers discussing this problem. The modified minimal 
subtraction ($\overline{MS}$) scheme \cite{c14,c15} is the simplest one, 
but it introduces the $\mu^2$-dependent terms which are very complicated to be 
dealt with. In the on-shell renormalization scheme, however, there isn't still 
an integrated CKM renormalization prescription. The early prescription 
\cite{c8} used the $SU_L(2)$ symmetry of SM to relate the CKM counterterm with the
fermion wave-function renormalization constants (WRC) \cite{c16}. Although it is a 
delicate and simple prescription, it reduces the physical amplitude involving quark
mixing gauge dependent\footnote{This is easy to be understood since the $SU_L(2)$ 
symmetry of SM has been broken by the Higgs mechanism} \cite{c17,c18,c19}. 
A remedial method of this prescription is to replace the on-shell fermion WRC in 
the CKM counterterms with the ones calculated at zero momentum \cite{c17}. 
Another remedial method \cite{c19} is to rearrange 
the off-diagonal quark WRC in a manner similar to the pinch technique \cite{c20}. 

In the following section we discuss a CKM renormalization prescription which, 
after some modification satisfies the three criterions. Next we
generalize this prescription to n-loop level, simultaneously keep the satisfaction
of the three criterions. In section 4 we discuss the relationship between
the unitarity of the bare CKM matrix and the gauge independence of the CKM 
counterterms. Lastly we give our conclusions.

\section{One-Loop Renormalization of CKM Matrix}

Different from the idea of Ref.\cite{c8}, another idea is to formulate the 
CKM renormalization prescription with reference to the case of zero mixing. This 
has been done in Ref.\cite{c21,c12} at one loop level. The main idea is to 
renormalize the transition amplitude of W gauge boson 
decaying into up-type and down-type quarks to make it equal to the amplitude of 
the same process which is in the case of no quark generation mixing. 
In order to elaborate this idea we firstly introduce the one loop decaying
amplitude of $W^{+}\rightarrow u_i \bar{d}_j$ \cite{c21}

\beqa \iar
  T_1\,=\,A_L[V_{ij}(F_L+\frac{\delta g}{g}+\half\delta Z_W +
  \half\delta \bar{Z}^{uL}_{ii}+\half\delta Z^{dL}_{jj})+
  \sum_{k\not=i}\half\delta \bar{Z}^{uL}_{ik}V_{kj}+
  \sum_{k\not=j}\half V_{ik}\delta Z^{dL}_{kj}+\delta V_{ij}]+ \\ \hspace{10mm}
  V_{ij}[A_R F_R+B_L G_L+B_R G_R]\,,
\ear \eeqa
with $g$ and $\delta g$ the $SU(2)$ coupling constant and its counterterm, 
$\delta Z_W$ the W boson WRC, $\delta\bar{Z}^{uL}$ and $\delta Z^{dL}$ the 
left-handed up-type and down-type quark's WRC \cite{c10,c22}, and

\beqa \iar
  A_L\,=\,\frac{g}{\sqrt{2}}\bar{u}_i(p_1){\xslash \varepsilon}\gamma_L 
  \nu_j(q-p_1)\,, \\
  B_L\,=\,\frac{g}{\sqrt{2}}\bar{u}_i(p_1)\frac{\varepsilon\cdot p_1}{M_W}
  \gamma_L \nu_j(q-p_1)\,.
\ear \eeqa
where $\varepsilon^{\mu}$ is the W boson polarization vector, $\gamma_L$ and 
$\gamma_R$ are the left-handed and right-handed chiral operators, and $M_W$ is the
W boson mass. Similarly, replacing $\gamma_L$ with $\gamma_R$ in the above 
equations we can define $A_R$ and $B_R$ respectively. $F_{L,R}$ and $G_{L,R}$ are
four form factors. The main idea of Ref.\cite{c21} is to choose the CKM counterterm
to make the amplitude $T_1$ similar as the amplitude of W boson decaying into 
leptons, which has no fermion generation mixing. This idea is reasonable since 
such a renormalized amplitude will be ultraviolet finite and gauge independent 
\cite{c21}. After introducing the proper CKM counterterm the decaying amplitude of 
$W^{+}\rightarrow u_i \bar{d}_j$ will be changed to such form \cite{c21}:

$$ T_1\,=\,V_{ij}[A_L(F_L+\frac{\delta g}{g}+\half\delta Z_W +
   \half\delta \bar{Z}^{uL}_{ii[l]}+\half\delta Z^{dL}_{jj[l]})+
   A_R F_R+B_L G_L+B_R G_R] $$
where the subscript "[l]" denotes the quantity is obtained by replacing CKM matrix
elements with unit matrix elements. So it is easy to obtain \cite{c21}

$$ \delta V_{ij}\,=\,-\half\sum_k[\delta \bar{Z}^{uL}_{ik}V_{kj}+
   V_{ik}\delta Z^{dL}_{kj}]+\half V_{ij}[\delta \bar{Z}^{uL}_{ii[l]}+
   \delta Z^{dL}_{jj[l]}] $$

But in fact such a CKM counterterm doesn't make the decaying amplitude $T_1$ 
ultraviolet finite when $i\not=j$. It is easy to calculate the ultraviolet terms
of $T_1$ using this CKM counterterm, as follows,

$$ T_1 |_{UV-divergence}\,=\,\frac{\alpha V_{ij} \Delta}{32\pi M_W^2 s_W^2}
   (m_{d,i}^2-m_{d,j}^2+m_{u,j}^2-m_{u,i}^2) $$
with $\alpha$ the fine structure constant, $s_W$ the sine of the weak mixing angle
$\theta_W$, $m_{d,i}$ and $m_{d,j}$ the down-type quark's masses, $m_{u,i}$ and 
$m_{u,j}$ the up-type quark's masses, and 
$\Delta=2/(D-4)+\gamma_E-\ln(4\pi)+\ln(M^2_W/ \mu^2)$ (D is the space-time 
dimensionality, $\gamma_E$ is the Euler's constant and $\mu$ is an arbitrary 
mass parameter). This result shows that when $i\not=j$ the decaying amplitude of 
$W^{+}\rightarrow u_i \bar{d}_j$ will be ultraviolet divergent. We argue that the 
origination of this error comes from our one-sided knowledge about the difference
between the two cases of having and not having fermion generation mixing.
In the case of no fermion generation mixing only the same fermions as the
external-line fermions can appear at the fermion line that connects with the
external-line fermions in the Feynman diagrams of $W^{+}\rightarrow u_i \bar{d}_j$.
This is because if there is no generation mixing only
one generation fermions can appear at a fermion line in a Feynman diagram. 
The same restraint is also suitable for the fermion's mass counterterms and WRC
which appear at the fermion line that connects with the external-line fermions.
Of course no CKM matrix element appears at the fermion line that connects with 
the external-line fermions. On the other hand, if there has fermion generation 
mixing there is no such restraint. At one-loop level the difference 
between these two cases is at the fermion's WRC. Different from the result
of Eq.(24) in Ref.\cite{c21}, the amplitude $T_1$ will be the following form
in the case of no quark generation mixing,

\beq
  T_1\,=\,V_{ij}[A_L(F_L+\frac{\delta g}{g}+\half\delta Z_W +
  \half \delta \bar{Z}^{uL}_{ii[l]m_{d,i}\rightarrow m_{d,j}}+
  \half \delta Z^{dL}_{jj[l]m_{u,j}\rightarrow m_{u,i}})+A_R F_R+B_L G_L+B_R G_R] 
\eeq
So the CKM counterterm is obtained compared with Eq.(4) and Eq.(6)

\beq
  \delta V_{ij}\,=\,-\half\sum_k[\delta \bar{Z}^{uL}_{ik}V_{kj}+
  V_{ik}\delta Z^{dL}_{kj}]+\half V_{ij}[
  \delta \bar{Z}^{uL}_{ii[l]m_{d,i}\rightarrow m_{d,j}}+
  \delta Z^{dL}_{jj[l]m_{u,j}\rightarrow m_{u,i}}]
\eeq
Our calculation has shown this CKM counterterm is gauge independent and 
makes the physical amplitude $T_1$ convergent. 

As mentioned in section 1, the CKM renormalization prescription should keep the 
unitarity of the bare CKM matrix. Now we check this point. At one-loop level,
only four diagrams need to be considered when we calculate the CKM counterterm
in Eq.(7), as shown in Fig.1.

\begin{figure}[tbh]
\begin{center}
  \epsfig{file=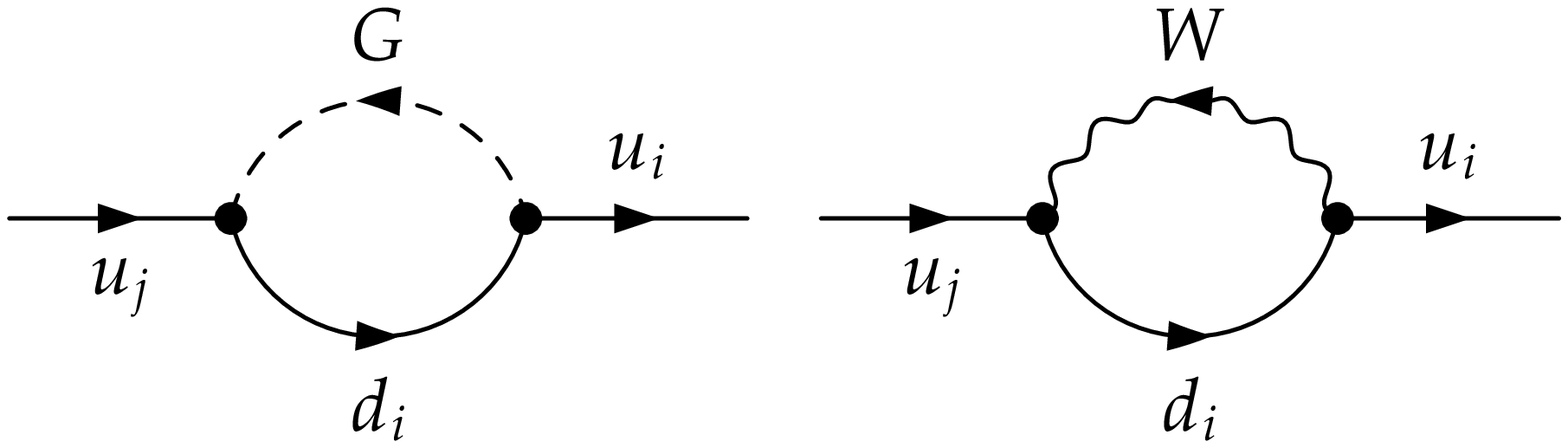, width=8cm} 
  \epsfig{file=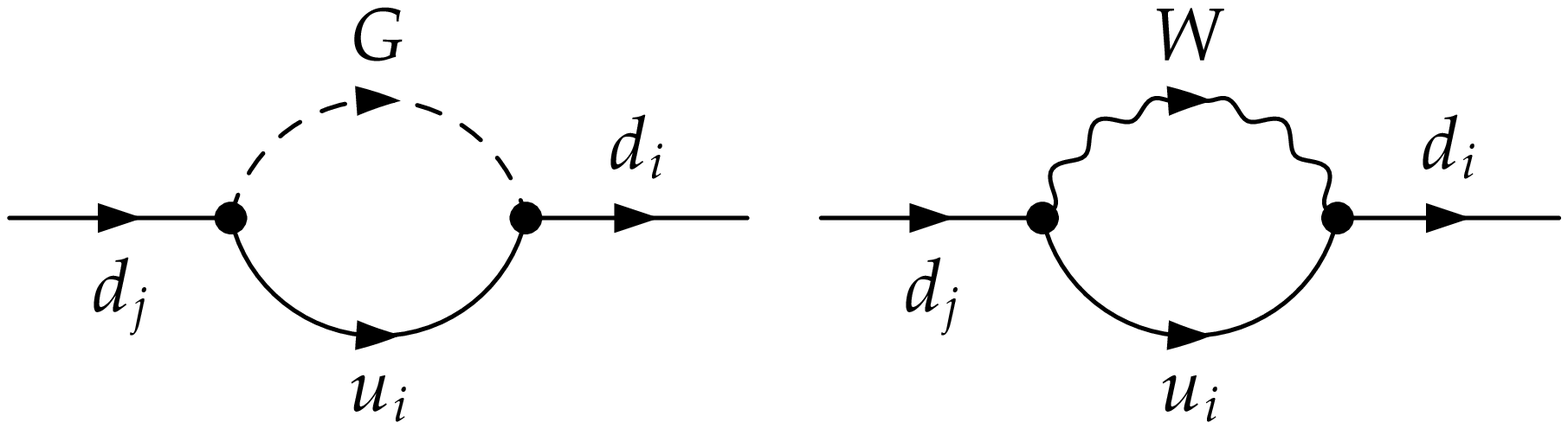, width=8cm}
  \caption{Quark's self-energy diagrams that contribute to the CKM counterterm
  in Eq.(7).}
\end{center} \end{figure}
We have used the software packages {\em FeynArts} \cite{c23} to draw the Feynman 
diagrams and generate the corresponding Feynman amplitudes, and used the software
packages {\em FeynCalc} \cite{c24} to calculate these Feynman amplitudes. We can 
obtain the analytical results of $\delta V_{ij}$ because the quark's self-energy 
functions are very simple. In order to check the unitary condition of Eq.(3) 
analytically, we use the Taylor's series: $(m^2_{quark}/M^2_W)^n$ to expand
$\delta V_{ij}$. The one and two order results are shown as follows:

\beqa \iar
  \hspace{-14mm}
  \delta V^{(1)}_{ij}\,=\,\frac{\alpha(6\Delta-11)}{128\pi M^2_W s^2_W}
  [-\frac{2\sum_{k,l\not=j}m_{d,j}m^2_{u,k}V_{il}V^{\ast}_{kl}V_{kj}}{m_{d,l}-
  m_{d,j}}+\frac{2\sum_{k,l}m_{d,j}m^2_{u,k}V_{il}V^{\ast}_{kl}V_{kj}}{m_{d,l}+
  m_{d,j}}-
  \frac{2\sum_{k\not=i,l}m_{u,i}m^2_{d,l}V_{il}V^{\ast}_{kl}V_{kj}}{m_{u,k}-
  m_{u,i}}+ \\  
  \frac{2\sum_{k,l}m_{u,i}m^2_{d,l}V_{il}V^{\ast}_{kl}V_{kj}}{m_{u,k}+
  m_{u,i}}+V_{ij}(\sum_k V_{ik}V^{\ast}_{ik}m^2_{d,k}+
  \sum_k V_{kj}V^{\ast}_{kj}m^2_{u,k}-2 m^2_{d,j}-2 m^2_{u,i})]\,.
\ear \eeqa

\beqa \iar
  \delta V^{(2)}_{ij}\,=\,\frac{\alpha}{64\pi M^4_W s^2_W}[V_{ij}
  (12 \ln\frac{m^2_{d,j}}{M^2_W} m^4_{d,j}+6 m^4_{d,j}+
  12 \ln\frac{m^2_{u,i}}{M^2_W} m^4_{u,i}+6 m^4_{u,i}+32 m^2_{d,j}m^2_{u,i}- 
  \\ \hspace{14mm}
  \sum_k V_{ik}V^{\ast}_{ik}(6 \ln\frac{m^2_{d,k}}{M^2_W} m^4_{d,k}+3 m^4_{d,k}+
  12 m^2_{u,i}m^2_{d,k})-\sum_k V_{kj}V^{\ast}_{kj}
  (6 \ln\frac{m^2_{u,k}}{M^2_W}m^4_{u,k}+3 m^4_{u,k}+12 m^2_{d,j}m^2_{u,k})) 
  \\ \hspace{14mm} 
  +\frac{2}{m_{d,l}-m_{d,j}}\sum_{k,l\not=j}m_{d,j}m^2_{u,k}(4 m^2_{d,j}+
  6 \ln\frac{m^2_{u,k}}{M^2_W} m^2_{u,k}+3 m^2_{u,k})V_{il}V^{\ast}_{kl}V_{kj} 
  \\ \hspace{14mm} -\frac{2}{m_{d,l}+m_{d,j}}\sum_{k,l}m_{d,j}m^2_{u,k}
  (4 m^2_{d,j}+6 \ln\frac{m^2_{u,k}}{M^2_W} m^2_{u,k}+3 m^2_{u,k})
  V_{il}V^{\ast}_{kl}V_{kj} \\ \hspace{14mm} 
  +\frac{2}{m_{u,k}-m_{u,i}}\sum_{k\not=i,l}m_{u,i}m^2_{d,l}(4 m^2_{u,i}+
  6 \ln\frac{m^2_{d,l}}{M^2_W} m^2_{d,l}+3 m^2_{d,l})V_{il}V^{\ast}_{kl}V_{kj} 
  \\ \hspace{14mm} 
  -\frac{2}{m_{u,k}+m_{u,i}}\sum_{k,l}m_{u,i}m^2_{d,l}(4 m^2_{u,i}+
  6 \ln\frac{m^2_{d,l}}{M^2_W} m^2_{d,l}+3 m^2_{d,l})V_{il}V^{\ast}_{kl}V_{kj}]\,.
\ear \eeqa
where the superscript "(1)" and "(2)" denote the one order and two order results
about the series $m^2_{quark}/M^2_W$. Here we have used the dimensional 
regularization \cite{c25} and $R_{\xi}$-gauge \cite{c26}. Replacing $\delta V$ with
$\delta V^{(1)}+\delta V^{(2)}$ in Eq.(3), we find at one-loop level they satisfy
the unitary condition.  

But when we consider the three order result of $\delta V$ about the series
$m^2_{quark}/M^2_W$, we find it doesn't satisfy the unitary condition, as shown 
below:

\beqa \iar
  \sum_k(\delta V^{(3)\ast}_{ki}V_{kj}+V^{\ast}_{ki}\delta V^{(3)}_{kj})\,=\,
  \frac{9\alpha}{128\pi M^6_W s^2_W}[\sum_k m^2_{u,k}(m^4_{d,i}-
  2 m^2_{d,j}m^2_{d,i}+m^4_{d,j}+m^2_{d,i}m^2_{u,k}+
  m^2_{d,j}m^2_{u,k})V^{\ast}_{ki}V_{kj} \\ \hspace{46mm} 
  -2\sum_{k,l,n}m^2_{u,k}m^2_{d,l}m^2_{u,n}V^{\ast}_{ki}V_{kl}V^{\ast}_{nl}V_{nj}]
  \,\not=\,0
\ear \eeqa
which shows that $\delta V$ doesn't comply with the unitary criterion. We can 
estimate the deviation of 
$\sum_k(\delta V^{\ast}_{ki}V_{kj}+V^{\ast}_{ki}\delta V_{kj})$ from 0. We argue 
that only the terms that contain the largest series of $m^2_t/M^2_W$ are 
important. Calculating to five order results of $\delta V$ about the series
$m^2_{quark}/M^2_W$, we find the largest deviation of 
$\delta V^{\dagger}V+V^{\dagger}\delta V$ from $0$ is 
proportional to $\alpha|V_{3i\not=3}|m^2_b m^8_t/(s_W^2 M^{10}_W) \sim 10^{-7}$, 
which is very small compared with the present measurement precision of the CKM 
matrix elements. Thus in actual calculations this deviation can be neglected.

Since $\delta V$ doesn't comply with the unitary criterion, Diener has put forward 
an amended prescription, to shift $\delta V_{ij}$ as \cite{c12}

\beq
  \delta\bar{V}_{ij}\,=\,\half(\delta V_{ij}-\sum_{k,l}V_{ik}\delta 
  V^{\dagger}_{kl}V_{lj})
\eeq
It is easy to check that $\delta\bar{V}$ satisfies the unitary criterion
at one-loop level. On the other hand, $\delta\bar{V}$ is gauge independent since
$\delta V$ is gauge independent, which makes the physical amplitude of  
$W^{+}\rightarrow u_i \bar{d}_j$ gauge independent \cite{c17,c12}. Because the 
ultraviolet divergence of $\delta V$ satisfies the unitary criterion \cite{c8}, 
it is easy to prove that $\delta \bar{V}$ has the same ultraviolet divergence 
as $\delta V$ \cite{c12}. So the new CKM counterterm $\delta \bar{V}$ satisfies
the three criterions mentioned in section 1 at one-loop level. 
 
\section{n-Loop Renormalization of CKM Matrix}

All of the mentioned prescriptions are only applied to one loop level. A suitable 
prescription for higher loop level is still not present. In view of the delicacy 
of Eq.(11), we want to follow the ideas of Ref.\cite{c21} and \cite{c12} to 
generalize them to be suitable for higher loop level. 

As we know when there is no quark generation mixing the amplitude of 
$W^{+}\rightarrow u_i \bar{d}_j$ doesn't need CKM renormalization. In other words,
in the case of no quark generation mixing the amplitude of 
$W^{+}\rightarrow u_i \bar{d}_j$ is gauge independent and ultraviolet finite after
introducing accurate physical parameter's counterterms except for the CKM 
renormalization counterterm.
Based on this point we want to choose the CKM counterterm to make the amplitude
of $W^{+}\rightarrow u_i \bar{d}_j$ equal to the amplitude of the same process that
is in the case of no quark generation mixing. Thus the CKM counterterm will be 
equal to the difference between the two amplitudes of 
$W^{+}\rightarrow u_i \bar{d}_j$ in the two different cases. Since the reason of 
introducing CKM counterterm is because of the existence of this difference, it is 
reasonable to make the CKM counterterm represents this difference.

Now our task is to find this difference. Although at an arbitrary loop level 
the Feynman diagrams of $W^{+}\rightarrow u_i \bar{d}_j$ are very complex, 
it is still very clearly that the difference between the two cases of having quark
generation mixing and zero-mixing (without quark generation mixing) 
only occurs at the fermion line that connects with the fermion external-lines.
According the discussion in section 2, the zero-mixing amplitude of 
$W^{+}\rightarrow u_i \bar{d}_j$ can be obtained by modifying the amplitude of 
$W^{+}\rightarrow u_i \bar{d}_j$ as

\begin{enumerate}
\item Do such treatment only at the fermion line that connects with the fermion 
external-lines: change the CKM matrix elements to unit matrix elements and CKM 
counterterms to zero; if there have odd number CKM matrix elements, leave a CKM 
matrix element unchanged (because the unit matrix element is a $\delta$ function,
the remaining CKM matrix element must be $V_{ij}$); then do such replacement:
$m_{u,j}\rightarrow m_{u,i}, m_{d,i}\rightarrow m_{d,j}$.
\item After the first step, change the fermion's mass counterterms and WRC 
appearing at the fermion line which connects with the fermion external-lines 
by the first step. 
\end{enumerate}

In order to determine the n-loop level CKM counterterm $\delta V_n$ we construct 
the n-loop level amplitude of $W^{+}\rightarrow u_i \bar{d}_j$ as follows 
(where only the n-loop level counterterms are listed for convenience) 

\beq
  T_n\,=\,A_L[F_{Ln}+V_{ij}(\frac{\delta g_n}{g}+\half\delta Z_{Wn})+
  \half\delta \bar{Z}^{uLn}V+\half V\delta Z^{dLn}+\delta V_n]+
  A_R F_{Rn}+B_L G_{Ln}+B_R G_{Rn}
\eeq
where the added denotation "n" represents the n-loop level result. According to the
above discussion we require this amplitude equal to the amplitude of the same 
process in the case of zero-mixing:

\beq
  T_n\,=\,A_L[F_{Ln[l]}+V_{ij}(\frac{\delta g_n}{g}+\half\delta Z_{Wn}+
  \half\delta \bar{Z}^{uLn}_{ii[l]}+\half\delta Z^{dLn}_{jj[l]})]+
  A_R F_{Rn}+B_L G_{Ln}+B_R G_{Rn}
\eeq
where the footnote "[l]" represents the new meaning: changing the quantity 
according to the two steps we have listed. So we can obtain the n-loop CKM 
counterterm by comparing Eq.(12) and (13)

\beq
  \delta V_{n}\,=\,F_{Ln[l]}+\half V_{ij}(\delta \bar{Z}^{uLn}_{ii[l]}+
  \delta Z^{dLn}_{jj[l]})-F_{Ln}-\half(\delta\bar{Z}^{uLn}V+V\delta Z^{dLn})
\eeq
Here we argue that such CKM counterterm will contain the subdivergences when they
appear at the right hand side of Eq.(14).

We have known that such CKM counterterm doesn't satisfy the unitary criterion at 
one-loop level. It needs to be modified. Here we introduce a new set of 
quantities: $\delta\bar{V}_1,\cdot\cdot\cdot,\delta\bar{V}_n$, the real 
CKM counterterm which satisfy the unitary criterion to n-loop level. Our aim is to
construct $\delta\bar{V}_n$ through 
$\delta V_n,\delta V_{n-1},\cdot\cdot\cdot,\delta V_1$, or equivalently,
through $\delta V_n,\delta \bar{V}_{n-1},\cdot\cdot\cdot,\delta \bar{V}_1$.
Here we require that $\delta V_n$ is obtained by using 
$\delta\bar{V}_{n-1},\cdot\cdot\cdot,\delta\bar{V}_1$ as the lower loop level CKM
counterterms in Eq.(14). Now the unitary criterion of Eq.(3) becomes 

\beqa \iar
  \delta\bar{V}_1 V^{\dagger}+V \delta\bar{V}^{\dagger}_1\,=\,0\,,\\
  \delta\bar{V}_2 V^{\dagger}+V \delta\bar{V}^{\dagger}_2\,=\,-
  \delta\bar{V}_1 \delta\bar{V}^{\dagger}_1\,,\\
  \delta\bar{V}_3 V^{\dagger}+V \delta\bar{V}^{\dagger}_3\,=\,-
  \delta\bar{V}_1 \delta\bar{V}^{\dagger}_2-
  \delta\bar{V}_2 \delta\bar{V}^{\dagger}_1\,,\\
  \cdot\cdot\cdot\cdot\cdot\cdot \\
  \delta\bar{V}_n V^{\dagger}+V \delta\bar{V}^{\dagger}_n\,=\,-
  \delta\bar{V}_1 \delta\bar{V}^{\dagger}_{n-1}-
  \delta\bar{V}_2 \delta\bar{V}^{\dagger}_{n-2}\cdot\cdot\cdot-
  \delta\bar{V}_{n-2}\delta\bar{V}^{\dagger}_2-
  \delta\bar{V}_{n-1}\delta\bar{V}^{\dagger}_1\,,\\
  \cdot\cdot\cdot\cdot\cdot\cdot 
\ear \eeqa
In order to solve these equations, we introduce a set of symbols $B_n$

\beqa \iar
  B_0\,=\,0\,, \\
  B_n\,=\,\sum^{n-1}_{i=1} -\delta\bar{V}_i \delta\bar{V}^{\dagger}_{n-i}\,.
\ear \eeqa
Obviously $B_n$ satisfies

\beq
  B_n\,=\,B^{\dagger}_n
\eeq
Assuming that we have obtained the counterterms 
$\delta\bar{V}_1,\delta\bar{V}_2,\cdot\cdot\cdot,\delta\bar{V}_{n-1}$ and 
$\delta V_n$, the n-loop level CKM counterterm $\delta\bar{V}_n$ can be obtained
in this way

\beq
  \delta\bar{V}_n\,=\,\half(\delta V_n -V\delta V^{\dagger}_n V+B_n V)
\eeq
Using induction it is easy to see that such CKM counterterms will satisfy the
unitary criterion to n-loop level.

The next step is to test whether the new CKM counterterm 
$\delta\bar{V}_1+\delta\bar{V}_2+\cdot\cdot\cdot+\delta\bar{V}_n$ satisfies the
two other criterions: make the physical amplitudes convergent and gauge 
independent. We can use induction to prove this point. The one-loop level
result has been proven in section 2. We only need to prove that if 
$\delta\bar{V}_1+\cdot\cdot\cdot+\delta\bar{V}_{n-1}$ satisfies the two criterions
to $n-1$ loop level, $\delta\bar{V}_1+\cdot\cdot\cdot+\delta\bar{V}_n$ will 
satisfy the two criterions to n-loop level. To do so we only need to prove the 
divergent and gauge-dependent part of $\delta\bar{V}_n$ equal to the divergent 
and gauge-dependent part of $\delta V_n$, since the latter contains the exact 
n-loop level divergent and gauge-dependent terms. Based on the renormalizability 
and predictability of SM, we can predict that the divergent and gauge-dependent 
part of $\delta V_n$ must satisfy the unitary criterion at n-loop level,

\beq
  \delta V^{DG}_n V^{\dagger}+V\delta V^{DG\dagger}_n\,=\,B^{DG}_n
\eeq
where the superscript "DG" denotes the divergent or gauge dependent part of 
the quantity. This is because if not so the unitary condition of the bare CKM 
matrix will require the divergent and gauge dependent part of the real CKM 
counterterm different from $\delta V_n$ thus will reduce the physical amplitude
of $W^{+}\rightarrow u_i \bar{d}_j$ divergent and gauge dependent.
Using Eq.(18) and (19) we obtain

\beq
  (\delta\bar{V}^{DG}_n-\delta V^{DG}_n)V^{\dagger}\,=\,\half(B^{DG}_n-
  \delta V^{DG}_n V^{\dagger}-V\delta V^{DG\dagger}_n)\,=\,0
\eeq
This identity manifests that 

\beq
  \delta\bar{V}^{DG}_n\,=\,\delta V^{DG}_n
\eeq
i.e. $\delta\bar{V}_n$ contains the same divergent and gauge dependent terms as 
$\delta V_n$. So we have proven the CKM counterterm
$\delta\bar{V}_1+\delta\bar{V}_2+\cdot\cdot\cdot+\delta\bar{V}_n$ 
satisfies the three criterions to n-loop level.

Now we have obtained the suitable CKM counterterm to n-loop level. 
$\{\delta\bar{V}_1,\delta\bar{V}_2,\cdot\cdot\cdot,\delta\bar{V}_n,
\cdot\cdot\cdot\}$ constructs a series. Here we list the results of 
$\delta\bar{V}_1,\delta\bar{V}_2,\delta\bar{V}_3$ and $\delta\bar{V}_4$

\beqa \iar
  \delta\bar{V}_1\,=\,\half(\delta V_1 -V\delta V^{\dagger}_1 V)\,, \\
  \delta\bar{V}_2\,=\,\half(\delta V_2 -V\delta V^{\dagger}_2 V)+
  \frac{1}{8}(\delta V_1 V^{\dagger}\delta V_1+
  V\delta V^{\dagger}_1 V\delta V^{\dagger}_1 V-V\delta V^{\dagger}_1\delta V_1-
  \delta V_1\delta V^{\dagger}_1 V)\,, \\
  \delta\bar{V}_3\,=\,\half(\delta V_3 -V\delta V^{\dagger}_3 V)+
  \frac{1}{8}(\delta V_1 V^{\dagger}\delta V_2+\delta V_2 V^{\dagger}\delta V_1+
  V\delta V^{\dagger}_1 V\delta V^{\dagger}_2 V+
  V\delta V^{\dagger}_2 V\delta V^{\dagger}_1 V- \\ \hspace{11mm} 
  V\delta V^{\dagger}_1\delta V_2-\delta V_2 \delta V^{\dagger}_1 V-
  V\delta V^{\dagger}_2\delta V_1-\delta V_1\delta V^{\dagger}_2 V)\,, \\
  \delta\bar{V}_4\,=\,\half(\delta V_4 -V\delta V^{\dagger}_4 V)+
  \frac{1}{8}(\delta V_1 V^{\dagger}\delta V_3+\delta V_3 V^{\dagger}\delta V_1+
  \delta V_2 V^{\dagger}\delta V_2+V\delta V^{\dagger}_1 V\delta V^{\dagger}_3 V+
  V\delta V^{\dagger}_3 V\delta V^{\dagger}_1 V+ \\ \hspace{11mm}
  V\delta V^{\dagger}_2 V\delta V^{\dagger}_2 V-V\delta V^{\dagger}_1\delta V_3-
  \delta V_3\delta V^{\dagger}_1 V-V\delta V^{\dagger}_2\delta V_2-
  \delta V_2\delta V^{\dagger}_2 V-V\delta V^{\dagger}_3\delta V_1-
  \delta V_1\delta V^{\dagger}_3 V)+ \\ \hspace{11mm}
  \frac{1}{128}(\delta V_1 V^{\dagger}\delta V_1\delta V^{\dagger}_1\delta V_1+
  \delta V_1\delta V^{\dagger}_1\delta V_1 V^{\dagger}\delta V_1+
  V\delta V^{\dagger}_1 V\delta V^{\dagger}_1 V\delta V^{\dagger}_1\delta V_1+
  V\delta V^{\dagger}_1 V\delta V^{\dagger}_1 \delta V_1\delta V_1^{\dagger}V+ 
  \\ \hspace{11mm}
  V\delta V^{\dagger}_1 \delta V_1 V^{\dagger}\delta V_1 V^{\dagger}\delta V_1+
  V\delta V^{\dagger}_1 \delta V_1 \delta V_1^{\dagger}V \delta V_1^{\dagger}V+
  \delta V_1 V^{\dagger}\delta V_1 V^{\dagger}\delta V_1\delta V_1^{\dagger}V+
  \delta V_1 \delta V_1^{\dagger}V\delta V_1^{\dagger}V\delta V_1^{\dagger}V- 
  \\ \hspace{11mm}
  V\delta V^{\dagger}_1\delta V_1\delta V^{\dagger}_1\delta V_1-
  \delta V_1\delta V^{\dagger}_1 V\delta V^{\dagger}_1\delta V_1-
  \delta V_1\delta V^{\dagger}_1\delta V_1\delta V^{\dagger}_1 V-
  V\delta V^{\dagger}_1 V\delta V^{\dagger}_1\delta V_1 V^{\dagger}\delta V_1-
  V\delta V_1^{\dagger}\delta V_1 V^{\dagger}\delta V_1\delta V_1^{\dagger}V 
  \\ \hspace{11mm} 
  -\delta V_1 V^{\dagger}\delta V_1 V^{\dagger}\delta V_1 V^{\dagger}\delta V_1-
  \delta V_1 V^{\dagger}\delta V_1\delta V_1^{\dagger}V\delta V_1^{\dagger}V-
  V\delta V_1^{\dagger}V\delta V_1^{\dagger}V\delta V_1^{\dagger}V\delta 
  V_1^{\dagger}V)\,.
\ear \eeqa
We guess our CKM renormalization prescription doesn't break the present symmetries 
of SM, e.g. Ward-Takahashi identity, since it only changes the value of CKM matrix
elements from $V^0_{ij}$ to $V_{ij}+\delta\bar{V}_{ij}$. 

\section{Relationship between the Unitarity and Gauge Independence of CKM Matrix}

It has been proven using Nielsen identities \cite{c27} that any physical 
parameter's counterterm must be gauge independent \cite{c13,c28}. The CKM matrix
elements are physical parameters so their counterterms must be gauge independent.
At one-loop level this conclusion has been proven \cite{c17}. When considering
higher loop level case a concrete problem will arise that whether the choice of 
the values of the lower loop level CKM counterterms will affect the gauge 
independence of the higher loop level CKM counterterms? As we know at one-loop 
level one can choose the gauge-independent convergent part of the CKM counterterm
freely. Will the different choices change the gauge independence of the CKM 
counterterm at two loop level? In order to clarify this problem we express the 
amplitude of $W^{+}\rightarrow u_i \bar{d}_j$ as

\beq
  T(V^0)\,=\,T(V+\delta V)\,=\,T(V)+T^{\prime}(V)\delta V+
  \half T^{\prime\prime}(V)(\delta V)^2+\cdot\cdot\cdot
\eeq
where the superscript $\prime$ denotes the partial derivative with respect to CKM 
matrix of the quantity. To two loop level, this equation becomes

\beq
  T_2(V^0)\,=\,T_2(V)+T^{\prime}_1(V)\delta V_1+\delta V_2 A_L
\eeq
with $T_2(V)$ and $T_1(V)$ the 2-loop and 1-loop amplitudes of 
$W^{+}\rightarrow u_i\bar{d}_j$ which don't contain CKM counterterms. In order to
find the effect of the choice of $\delta V_1$ on the gauge independence of 
$\delta V_2$ we need to calculate $T^{\prime}_1(V) \delta V_1$ analytically. 
From Eq.(4), since $F_R$ and $G_{L,R}$ are gauge independent and don't contain 
CKM matrix element, only the terms in the first bracket of Eq.(4) need to be 
considered. Using the fact that the terms in the first bracket of Eq.(4) is gauge 
independent \cite{c17}, we have

\beqa \iar
  T^{\prime}_1(V)\delta V_1 |_{\xi}=[-\frac{\delta V_{ij}}{2 V_{ij}}
  (\sum_{k\not=i}\delta \bar{Z}^{uL}_{ik}V_{kj}+
  \sum_{k\not=j}V_{ik}\delta Z^{dL}_{kj})+\frac{V_{ij}}{2}\sum_{k,l}
  (\frac{2}{g}\frac{d(\delta g)}{d V_{kl}}+\frac{d(\delta Z_W)}{d V_{kl}}+
  \frac{d(\delta \bar{Z}^{uL}_{ii})}{d V_{kl}}+
  \frac{d(\delta Z^{dL}_{jj})}{d V_{kl}})\delta V_{kl}+ \\ \hspace{21mm} 
  \half(\sum_{k\not=i}\delta \bar{Z}^{uL}_{ik}\delta V_{kj}+
  \sum_{l,m,k\not=i}\frac{d(\delta \bar{Z}^{uL}_{ik})}{d V_{lm}}\delta V_{lm}V_{kj}
  +\sum_{k\not=j}\delta V_{ik}\delta Z^{dL}_{kj}+
  \sum_{l,m,k\not=j}V_{ik}\frac{d(\delta Z^{dL}_{kj})}{d V_{lm}}\delta V_{lm})]A_L
\ear \eeqa
where the subscript "1" of $\delta V_1$ has been omitted and the subscript "$\xi$" 
on the left hand side of this equation denotes the gauge dependent part of 
the quantity. Omitting the imaginary parts of the quark's self energies (because
they are gauge independent), we obtain

\beqa \iar
  T^{\prime}_1(V)\delta V_1 |_{\xi}=\frac{\alpha A_L}{32\pi M^2_W s^2_W m^2_{d,j}}
  \sum_{k,l}(\delta V_{il}V^{\ast}_{kl}+V_{il}\delta V^{\ast}_{kl})V_{kj}[-
  \xi^2_W M_W^4\ln\frac{m_{u,k}^2}{M_W^2}+
  2\xi_W m_{d,j}^2 M_W^2\ln\frac{m_{u,k}^2}{M_W^2}-m_{u,k}^4\ln\xi_W 
  \\ \hspace{21mm} 
  +2 m_{d,j}^2 m_{u,k}^2\ln\xi_W-2\sqrt{-\xi_W^2 M_W^4+2\xi_W m_{d,j}^2 M_W^2
  +2\xi_W m_{u,k}^2 M_W^2-m_{d,j}^4-m_{u,k}^4+2 m_{d,j}^2 m_{u,k}^2} 
  \\ \hspace{21mm}
  (\xi_W M_W^2-m_{d,j}^2+m_{u,k}^2)\arctan\frac{\sqrt{-\xi_W^2 M_W^4+
  2\xi_W m_{d,j}^2 M_W^2+2\xi_W m_{u,k}^2 M_W^2-m_{d,j}^4-m_{u,k}^4+
  2 m_{d,j}^2 m_{u,k}^2}}{-\xi_W M_W^2+m_{d,j}^2-m_{u,k}^2}]+ 
  \\ \hspace{21mm} 
  \frac{\alpha A_L}{32\pi M^2_W s^2_W m^2_{u,i}}\sum_{k,l}
  (V^{\ast}_{lk}\delta V_{lj}+\delta V^{\ast}_{lk}V_{lj})V_{ik}[-
  \xi^2_W M_W^4\ln\frac{m_{d,k}^2}{M_W^2}+
  2\xi_W m_{u,i}^2 M_W^2\ln\frac{m_{d,k}^2}{M_W^2}-m_{d,k}^4\ln\xi_W
  \\ \hspace{21mm}
  +2 m_{u,i}^2 m_{d,k}^2\ln\xi_W -2\sqrt{-\xi_W^2 M_W^4+2\xi_W m_{u,i}^2 M_W^2
  +2\xi_W m_{d,k}^2 M_W^2-m_{u,i}^4-m_{d,k}^4+2 m_{u,i}^2 m_{d,k}^2} 
  \\ \hspace{21mm}
  (\xi_W M_W^2-m_{u,i}^2+m_{d,k}^2)\arctan\frac{\sqrt{-\xi_W^2 M_W^4+
  2\xi_W m_{u,i}^2 M_W^2+2\xi_W m_{d,k}^2 M_W^2-m_{u,i}^4-m_{d,k}^4+
  2 m_{u,i}^2 m_{d,k}^2}}{-\xi_W M_W^2+m_{u,i}^2-m_{d,k}^2}]
\ear \eeqa
with $\xi_W$ the W boson gauge parameter. It can be seen that if $\delta V_1$ 
satisfies the unitary criterion, the gauge dependent part
of $T^{\prime}_1(V)\delta V_1$ will be equal to zero. On the other hand, from
Eq.(24) we can see that the gauge dependent part of $\delta V_2$ is determined by 
the gauge dependent part of $T_2(V)$ and $T^{\prime}_1(V)\delta V_1$ since
$T_2(V_0)$ is gauge independent. So we know as long as $\delta V_1$ satisfies 
the unitary criterion, no matter how to change the value of $\delta V_1$ will not
affect the gauge independence of $\delta V_2$. On the contrary, if $\delta V_1$ 
doesn't satisfy the unitary criterion, changing the value of $\delta V_1$ will
change the gauge dependent part of $\delta V_2$, thus will make $\delta V_2$
gauge dependent. Therefore we can draw a conclusion that only if keep the unitarity
of the bare CKM matrix in the CKM matrix renormalization prescription 
the renormalized CKM matrix and its counterterm will be gauge independent.

\section{Conclusion}

In summary, we have investigated the present CKM matrix renormalization 
prescriptions and found all of them are only suitable for one loop level. 
In this paper, we have checked the prescription in Ref.\cite{c21} and found
it doesn't satisfy the unitary criterion of the bare CKM matrix. There is also 
an error in this prescription. The correct one-loop CKM counterterm should
be the form of Eq.(7). Then we generalize the prescription in Ref.\cite{c21} and
\cite{c12} to make it suitable for any loop level and comply with the 
unitary criterion of the bare CKM matrix. The concrete results are shown in 
Eq.(18) and Eqs.(22). Our prescription also makes the amplitude of an arbitrary
physical process involving quark mixing convergent and gauge independent, 
as required. Lastly We point out that only if the CKM renormalization prescription
keeps the unitarity of the bare CKM matrix the renormalized CKM matrix and its
counterterm will be gauge independent.

\vspace{5mm} {\bf \Large Acknowledgments} \vspace{2mm} 

The author thanks professor Xiao-Yuan Li for his useful guidance and Dr. Hu 
qingyuan for his sincerely help (in my life).

\end{document}